\documentclass[reviewcopy,11pt]{elsart}
\usepackage[dvips]{graphicx}
\usepackage{epsfig}
\usepackage{amsmath}
\usepackage{amssymb}

\begin{document}
\begin{frontmatter}


\title{On the Potential of the Excluded Volume and Auto-Correlation as
   Neuromorphometric Descriptors}

\author[label1]{L. da F. Costa, M. S. Barbosa\corauthref{cor1}}
\author[label2]{ and V. Coupez}
\corauth[cor1]{FAX: 55 16 273 9879, {\it email:}  marconi@if.sc.usp.br}
\address[label1]{Cybernetic Vision Research Group
Instituto de F\'{\i}sica de S\~ao Carlos
University of S\~ao Paulo
13560-970  S\~ao Carlos, SP, Brazil}

\address[label2]{Polytechnic Institute of Orl\'eans,
ESPEO. 12, rue de Blois. BP 6744 45067 Orl\'eans, Cedex 2, France}

\begin{abstract}

 This work investigates at what degree two neuromorphometric
 measurements, namely the autocorrelation and the excluded volume of
 a neuronal cell can influence the characterization and
 classification of such a type of cells.  While the autocorrelation
 function presents good potential for quantifying the
 dendrite-dendrite connectivity of cells in mosaic tilings, the
 excluded volume, i.e. the amount of the surround space which is
 geometrically not accessible to an axon or dendrite, provides a
 complementary characterization of the cell connectivity. The
 potential of such approaches is illustrated with
 respect to real neuronal cells.

\end{abstract}
\end{frontmatter}

\section{Introduction}

Neuroanatomy is, in many aspects, a key factor influencing the overall nervous
system performance, regarding health and adaptation, underlined by the spatial
characteristics of neural networks and by the geometry of their building
blocks, the neuronal cells.  The function of such assemblage of cells, which
integrates synaptic propagation of action potentials, has long been related to
the shape of the individual cell and specific dendritic
morphology~\cite{cajal:1990}.  The study of morphological characteristics of
neuronal cells, although overshadowed by the flood of advances in
neurophysiological explorations along the last decades, was resumed in the
seventies as an approach in which computer science methods provide the means
for quantifying, in an objective way, morphological subtleties that would
otherwise go unnoticed by the human eye ~\cite{wassle:1974}. Examples of
geometrical features commonly used in computational neuromorphology include
the influence area, number of dendritic segments and bending energy
(see~\cite{LU020} for an extensive list) and as well as standard and
multi-scale fractal analysis
\cite{MOR89,Jelinek:2001b,Jelinek:2001c,Coelho:1996,LU101,smith:1996}. It has
also been shown more recently that neuronal cell characterization can be
efficiently performed with basis on multi-scale shape functionals derived from
integral geometry and applied to a database containing two classes of cat
ganglion cells~\cite{barbosa:2003a}.

In the current work we extend geometrical features that are not only useful
for neuromorphic characterization, but which are also more closely related to
the functionality of the network, with emphasis on their abilities at making
connections as highlighted in~\cite{shefi:2002,morita:2001,tadeu:2002}.  In
particular, we explore and assess the use of the autocorrelation function and
excluded volume of the shape of a neuronal cell as meaningful measurements to
be used in cell characterization, classification and simulation.  The reader
is reported to~\cite{rodieck:1991,cook:1996,wassle:1981} for pioneering works
related to these concepts.  It is shown here that the autocorrelation function
quantifies in a precise fashion the potential of the cell for
dendrite-dendrite connections, a situation that is especially relevant for
neuronal systems involving electrical synapses.  At the same time, the
excluded volume is verified to provide additional information about the cell
potential for connectivity, but now from the complementary perspective of
unreachable portions of the arborizations.

The article starts by presenting in Section 2 the suggested
measurements and by describing their respective application to a few
real ganglionary cells. Section 3 presents our results for an extended
database of neuronal cells containing 50 cells, showing the
discriminative potential of the proposed methodology. Section 4 is
dedicated to concluding comments.

\section{Shape Descriptors}

This section describes the two considered neuronal features, namely
the two-dimensional autocorrelation function and the excluded
volume. These measures have in common the conceptual advantage of
being related to the connectivity potential of the cell in an actual
simulation of a morphological neuronal network. Although the excluded
volume and the autocorrelation function are not multi-scale
functionals, the former situation demands more intense computational
efforts when compared to the latter which is faster as its
implementation can be done by using properties of Fourier analysis.

\subsection{Autocorrelation Function}

Given a function $\varphi(x,y)$, its autocorrelation $\alpha(x,y)$ is
defined as in Equation~\ref{eq:autoc}, which can be effectively
evaluated by using the Fourier transform (Equation~\ref{eq:four_ac}).

\begin{eqnarray}
\alpha(x,y)=\int_{-\infty}^{\infty}\int_{-\infty}^{\infty}
 \varphi(r,s) \varphi(r+x,s+y) dr ds \label{eq:autoc} \\
 \alpha(x,y)=\mathcal{F}^{-1}( \mathcal{F} (\varphi(x,y)) \mathcal{F}(\varphi(x,y)^*)
  \label{eq:four_ac}
\end{eqnarray}

\begin{figure}[htb]
\begin{center}
\includegraphics[scale=.5,angle=0]{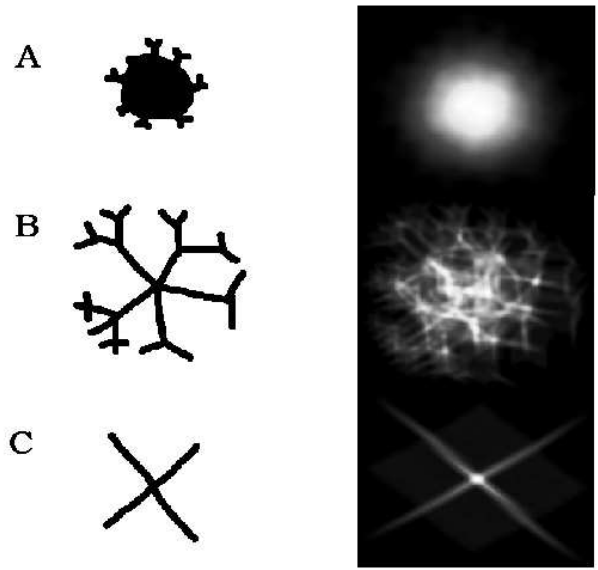}
\caption{The autocorrelation function for prototypical different cells.}
\label{fig:try}
\end{center}
\end{figure}

\begin{figure}[htb]
\begin{center}
\includegraphics[scale=.5,angle=0]{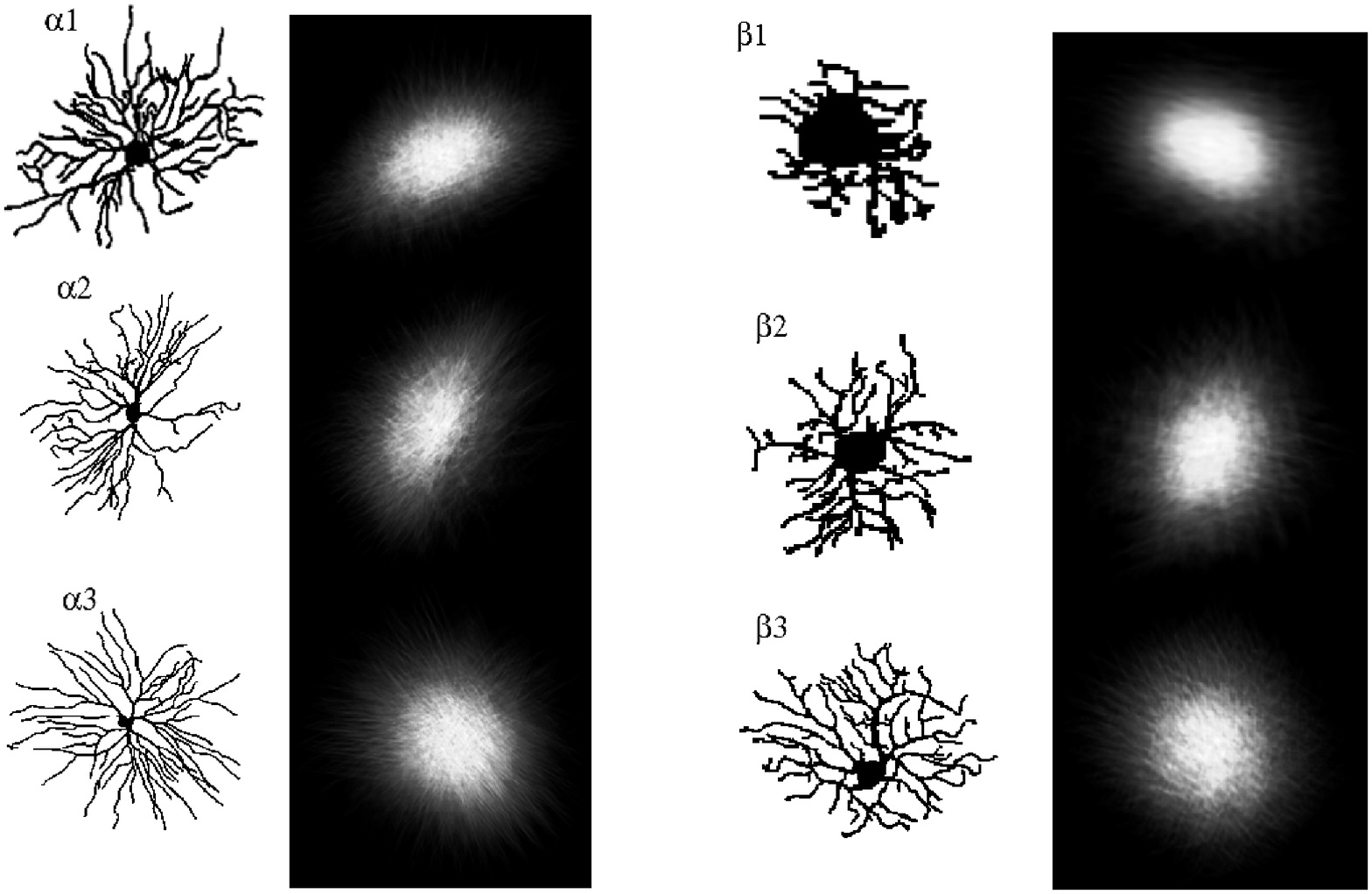}
\caption{The autocorrelation function of typical ganglion cells
 of the domestic cat retina from two phyisiological class. The alpha and beta
 types are shown at the left and right-hand sides, respectively. The neuronal
 cell images were adapted with permission from \cite{wassle:1974}.}
\label{fig:auto}
\end{center}
\end{figure}

Let the spatially sampled neuronal cell morphology (dendrites and
soma) be represented in terms of a binary image, with each cell point
represented as $1$ and taking the center of mass of the respective
soma (i.e. cell body) as the coordinate system origin.  The
autocorrelation of this shape at a specific point $(x,y)$,
i.e. $\alpha(x,y)$ corresponds to the number of intersection points
between two versions of the cell with relative displacement of
$(x,y)$.  Therefore, the autocorrelation function quantifies the cell
potential for dendrite-dendrite connections with respect to copies of
itself placed at all possible relative positions, see
Figure~\ref{fig:try} for examples of the autocorrelation for few
prototypical shapes.  Figure~\ref{fig:auto} illustrates the
autocorrelation function for a series of cat ganglion cells.  It is
important to observe that the information provided by the
autocorrelation analysis of the cell morphology represents a second
order analysis of the cell geometry, which could not be otherwise
obtained by more traditional measurements directly focusing the cell,
such as its area, perimeter, and so on.

The autocorrelation function is immediately related to the two
following biological questions: (a) in a relatively homogeneous
morphometric network (i.e. composed by similar copies of the same
cell) is there a preferential orientation where the cell presents
highest or lowest potential for connection? (b) the same question
regarding radial distance. Given a neuronal cell and its respective
two-dimensional autocorrelation function, it is therefore interesting
to derive some particularly meaningful one-dimensional signatures such
as the functions defined in Equations~\ref{eq:f_ang}
and~\ref{eq:f_rad}.

\begin{eqnarray}
 f(\theta) = \int_{r \in L(\theta)} \alpha(x,y) dr \label{eq:f_ang}  \\
 g(r) = \int_{\theta \in C(r)} \alpha(x,y) d \theta  \label{eq:f_rad}
\end{eqnarray}

where $L(\theta)$ is the line segment starting at the coordinate
system origin and making an angle of $\theta$ with the $x$-axis,
$\theta \in [0,2 \pi)$, and $C(r)$ is the circle of radius $r$
centered at the coordinate system origin.  While the function
$f(\theta)$ can be understood as the potential of the cell to make
dendrite-dendrite connections in terms of the angle $\theta$ made with
the $x$-axis, the function $g(r)$ gives the potential for connections
in terms of the distance $r$ from the soma center of mass.

\subsection{Excluded Volume}

The excluded volume for a given object is usually defined as the
volume surrounding and including a given object, which is excluded to
another object as it is placed at varying relative
positions~\cite{santalo:1976,onsager:1949}. A similar definition of
the excluded area holds in 2-D. Although we henceforth refer to this
concept as excluded "volume", it is important to keep in mind that it
actually refers to an area measurement taking into account the
topological features of the cell.  At the same time, 3D
generalizations are also possible and interesting and will be
considered in future works. This functional is always defined for a
pair of objects.

The {\it excluded volume} terminology comes from the statistical
mechanics of non-ideal gases, where this functional arises in the
leading order concentration expansion ({\it virial expansion} ) for
the pressure in the case of gas particles that repel each other with a
hard-core volume exclusion~\cite{isihara:1950}.  This idea is
frequently employed also in systems that share the same physical
constraints such as colloidal solutions in polymer sciences.

The general expression for the excluded volume $V_{ex}$ of two convex
objects, involving their surface area $F_i$ and mean radius of
curvature of each $R_i$ and their individual volumes, $V_i$, is given
by (see~\cite{isihara:1950}):

\begin{equation}
<V_{ex}>=V_1+V_2+(A_1R_1 +A_2R_2)/(4\pi)
\end{equation}

\begin{figure}[htb]
\begin{center}
\includegraphics[scale=.5,angle=0]{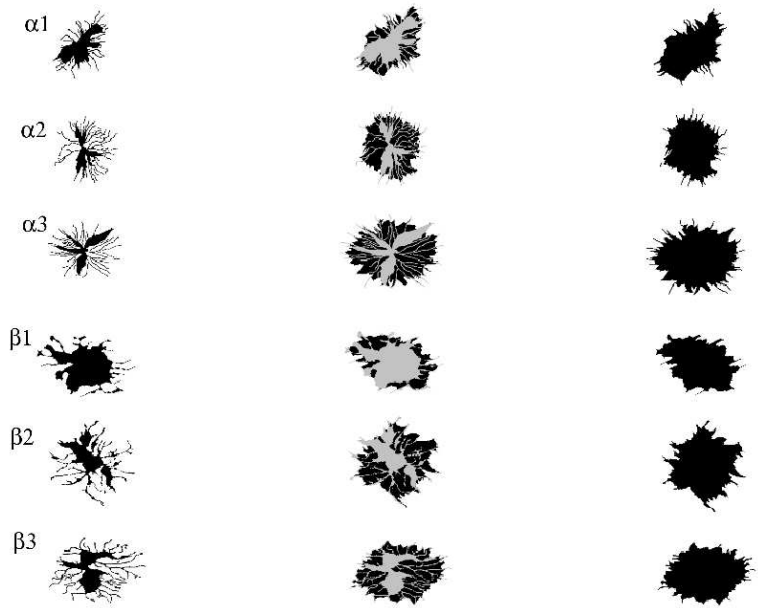}
\caption{The excluded volume for examples of real ganglion cells
 of the domestic cat retina. The neuronal cell images were adapted with
 permission from \cite{wassle:1974}.}
\label{fig:exvol}
\end{center}
\end{figure}

Here we evaluate a computational variant of the excluded volume
concept. More specifically, our algorithm tries to fit a copy of the
cell shape at every relative position on an orthogonal grid, while
checking for intersections. Whenever no intersection is found, the
respective region below the image is considered accessible, and the
excluded volume function at that specific point is marked as $0$,
otherwise it is marked as $1$.  Figure~\ref{fig:exvol} illustrates the
excluded volume function obtained for several different cells.

It stems from the above definition that all convex shapes have a null
excluded volume. For an intricate geometry such as neuronal cell forms
this functional can be understood as an indication of the boundaries
of connectable portions of the cell. From the functional point of
view, all dendrites which are trapped inside an excluded volume region
will not actually establish connections with other cells, compromising
the performance of a network made with copies of that same cell. This
fact is being further investigated as a part of an ongoing research on
morphological Hopfield networks~\cite{barbosa:2003b}.
 
\section{Results and Discussion}

In this work we use a database of 53 images of ganglion cells from the
domestic cat retina of which 26 are of type alpha and the remaing 27
of type beta. These two types of neuronal cells can often be difficult
to be distiguished by human visual inspection.  Typical
cells from these classes are illustrated in Figures~\ref{fig:auto}
and~\ref{fig:exvol}.

\begin{figure}[htb]
\begin{center}
\includegraphics[scale=.4,angle=-90]{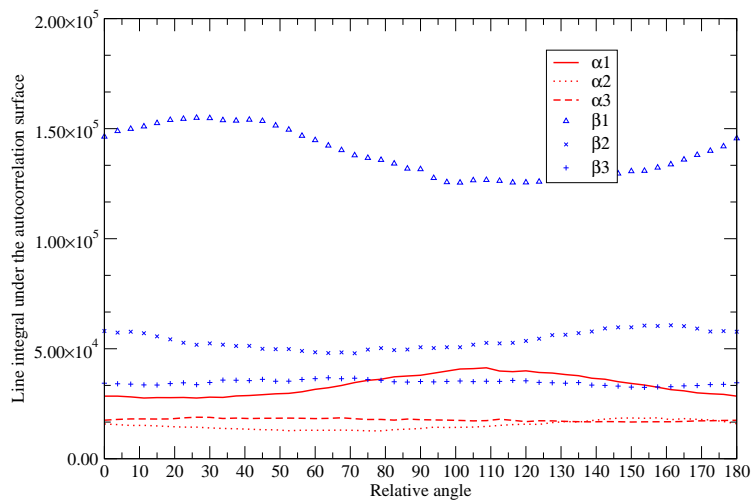}
\caption{The line integral of the correlation function in terms of the
 angle of a line passing through the maximum point of correlation. In this
 figure we show results for typical cells of the two physiological classes of
 cat ganglion cells shown in figure~\ref{fig:auto}.}
\label{fig:angle}
\end{center}
\end{figure}

\begin{figure}[htb]
\begin{center}
\includegraphics[scale=.4,angle=-90]{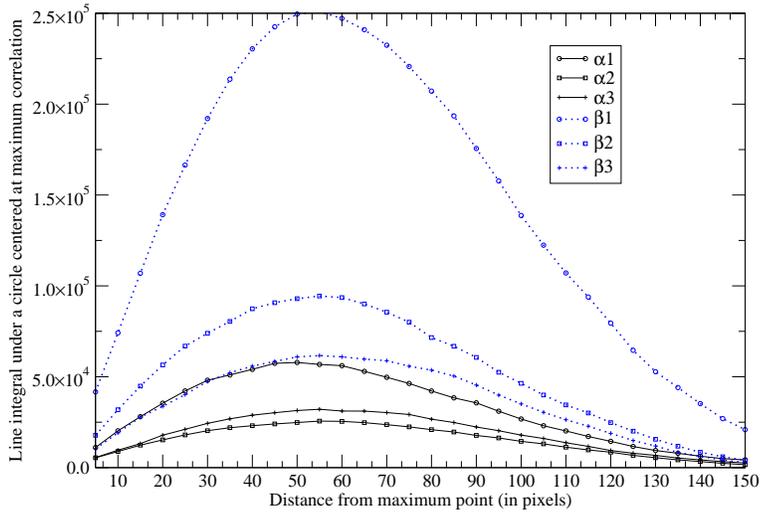}
\caption{The line integral of the correlation function in terms of the
 radius of a circle centered at the maximum point of correlation. In this
 figure we show results for the same typical cells of the domestic cat
 ganglion cells shown in figure~\ref{fig:auto}.}
\label{fig:radius}
\end{center}
\end{figure}

Figures~\ref{fig:angle} and~\ref{fig:radius} present the functions
$f(\theta)$ and $g(r)$ (Equations~\ref{eq:f_ang},~\ref{eq:f_rad}) for
the ganglion cells in Figure~\ref{fig:auto}.  These especially
representative 6 cells were chosen for illustrative purposes, as the
functions $f(\theta)$ and $g(r)$ would result too cluttered in case
all 53 cells were used.  It is clear from the first of these figures
that quite different curves are obtained for each neuronal class,
suggesting that the considered measurements have good discriminative
power.  On the other hand, the curves obtained for the autocorrelation
line integrals in terms of the radial distance, shown in
Figure~\ref{fig:radius}, are characterized by similar overall
behaviour, but different magnitudes.  Although such a result implies
less discriminative power for this measurement, it also represents an
interesting biological phenomenon on its own, indicating that the
considered cells share quite similar potential for connections as far
as the radial distance is concerned.  More specifically, such curves
are characterized by small values at both left and right-hand
extremities, taking a maximum value in its intermediate portion.
While the small values for small distances is expected since line
integrals for those cases cover less perimeter, the small values
obtained for large radial distances is a consequence of the
progressive termination of the dendritic extremities.  It remains to
be further investigated whether similar curves are obtained for
classes of neuronal cells other than ganglion retinal neurons.

\begin{figure}[htb]
\begin{center}
\includegraphics[scale=.4,angle=-90]{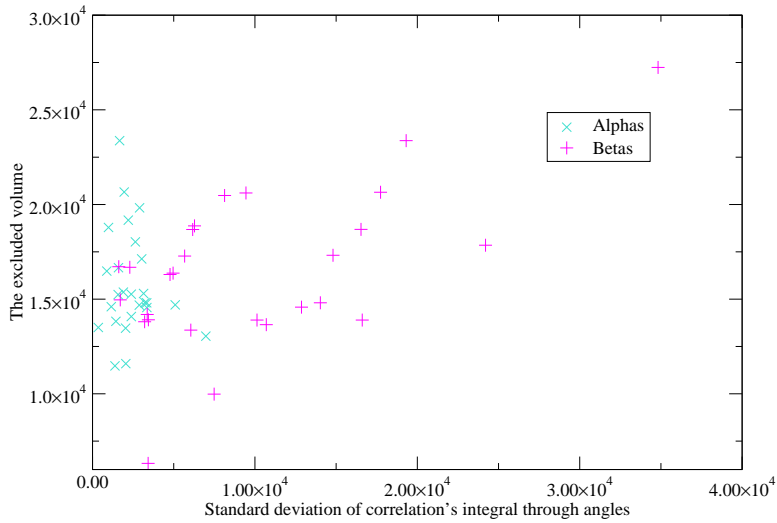}
\caption{A scatter plot, defined by measures of excluded volume and auto
 correlation, exhibiting significant clustering of two physiological classes
 of cat ganglion cells.}
\label{fig:scatter}
\end{center}
\end{figure}

Figure~\ref{fig:scatter} shows the phase space or scatter plot,
defined by the standard deviation of the correlation integral for
angles and the excluded volume, obtained for the considered ganglion
cells.  It is clear from this figure that the two classes of cells are
well-separated conforming to the discriminative potential of the
former of these features.  At the same time, this same figure
indicated that, at least for the considered data, the excluded volume
has less discriminative power, despite of its direct relationship
with the connectivity potential of a cell. This could be related to
the fact that the measure is not multiscale. Work to extend this
notion and to extract more information of this important concept is in
course.

\begin{table}[ht]
\begin{center}
\begin{tabular}{l|l|c|c|c|c}\hline\hline
        &  $\alpha$ &  $\beta$ &    Error & Posterior.Error & Stratified.Error\\\hline
$\alpha$ & 25 &  1 & 0.0384615 &     -0.1151105  &     -0.0437544\\\hline
$\beta$  &  5 & 22 & 0.2380952 &      0.2949808  &      0.2197885\\\hline
Overall  &    &    & 0.1382784 &      0.0899351  &      0.0880171\\\hline\hline
\end{tabular}
\end{center}
\caption{The result of classical discriminant analysis for the
measures considered in the scatterplot of Figure~\ref{fig:scatter}. From
this cross-validation table we observe a good discriminative
potential, most of it derived (see Figure~\ref{fig:scatter}) from the
autocorrelation functionals. \label{tab:cross} }
\end{table}

\vspace{2cm}

Table~\ref{tab:cross} shows the result of statistical discriminant
analysis \cite{McLachlan:1992} for all the 53 considered cells. This
cross-validation table shows clearly the potential of the proposed
methodology for proper cell classification. While the classification of
$\alpha$ cells was particularly effective with just one cell out of 26
misclassified, the identification of the beta cells was relatively
poorer with 5 out of the 27 cells being misclassified as alpha type.
This is in part a consequence of the fact that the cells are difficult
to be separated even by visual inspection.

Despite a few misclassified cells, the suggested measurements were
verified to be capable of discriminating between the 53 considered
cells.  However, it is important to bear in mind that the importance
of the suggested neuromorphometrical features goes beyond such an
ability for cell discrimination, in the sense that it reflects in a
quantitative way the potential of the investigated cells to establish
synaptic contacts, which is an all important property in biological
neuronal systems.  In addition, the excluded volume provides
additional information about the spatial delimitation of the neuronal
processes, which is also important in defining cell connections.  In
this way, the suggested methodology allows the prediction of
macroscopic properties of the neuronal system with basis on
local geometrical features of individual representative cells.

\section{Concluding Remarks}

This work presented a preliminary assessment of the potential for
neuromorphometric characterization of two features directly
related to the ability of a neuronal cell to make dendrite-dendrite
connections, which is immediately relevant for defining and
constraining the behaviour of neuronal systems.  In particular, we
have shown that the autocorrelation function leads to different
profiles for the cells in the considered ganglion cell database.  The
extracted features resulted particularly efficient for cell
classification. When taken on its own, the excluded volume led to
poorer discriminative potential. The continuation of this work should
include the correlation of the suggested measurements with functionals
expressing the performance of morphological neuronal
networks~\cite{barbosa:2003b} as well as other shape
functionals~\cite{barbosa:2003a,tadeu:2002}, 3D and multiscale
extensions of those measurements.  Another interesting perspective
that is being investigated is how the proposed measurements and
methodology, which implies neuronal networks composed by identical
copies of the same cell, can be extended according to a mean-field
approach, where a mean representation of the cell morphology is used
to analyse networks composed by cells with varying morphology.

\bibliographystyle{unsrt}
\bibliography{correl_icobi}

\end{document}